\documentclass[12pt,preprint]{aastex}
\usepackage{rotating}
\usepackage{epsfig}
\shorttitle{A Cluster of 1.3 cm Sources in OMC1 South}
\shortauthors{Zapata, Rodr\'\i guez, Kurtz, O'Dell, \& Ho}

\begin{document}

\title{A Cluster of 1.3 cm Continuum Sources in OMC1 South}

\author{Luis A. Zapata\altaffilmark{1,3}, 
Luis F. Rodr\'\i guez\altaffilmark{1}, Stanley, E. 
Kurtz\altaffilmark{1}, \\    
C. R. O'Dell\altaffilmark{2}, and Paul T. P. Ho
\altaffilmark{3}
}

\altaffiltext{1}{Centro de Radioastronom\'\i a y Astrof\'\i sica, 
UNAM, Apdo. Postal 3-72 (Xangari), 58089 Morelia, Michoac\'an, M\'exico}
\altaffiltext{2}{Department of Physics and Astronomy, Vanderbilt University, 
Box 1807-B, Nashville, TN 37235}
\altaffiltext{3}{Harvard-Smithsonian Center for Astrophysics, 60 Garden Street, 
Cambridge, MA 02138}

\email{lzapata@cfa.harvard.edu, l.rodriguez@astrosmo.unam.mx, s.kurtz@astrosmo.unam.mx, 
cr.odell@vanderbilt.edu, and pho@cfa.harvard.edu}
 
\begin{abstract}
We present sensitive 1.3 cm radio continuum observations of the region OMC1 South
(OMC-1S) in Orion using 
the Very Large Array in its B configuration. We detect eleven  
radio sources clustered in a $30{''} \times 30{''}$ region,
of which only three had been detected previously at
radio wavelengths in deep 
3.6 cm observations. 
The eight new radio sources are compact ($\theta_s \leq 0\rlap.{''}1$)
and we set lower limits to their spectral indices, $\alpha > 0.8 \pm 0.3$
(with $S_\nu \propto \nu^{\alpha}$),
that suggest that they may be optically-thick 
H~II regions. However, one of the new sources exhibits
significant circular polarization, indicating that gyrosynchrotron
emission with large positive spectral indices may be an
alternative explanation.
Furthermore, we find that
four other sources are associated with infrared sources
of low bolometric luminosity that cannot drive an H~II region. 
Finally, two of the sources previously detected 
at 3.6-cm are angularly resolved in the 1.3 cm image 
and their major axes have position angles that align well
with large scale outflows emanating from OMC-1S.
The radio source 143-353 has a major axis with a position angle consistent
with those of the HH~202 and HH~528 flows, while the radio source 134-411
has a major axis with a position angle consistent
with that of the low-velocity
molecular outflow associated with the far-infrared source FIR~4.

\end{abstract}  


\section{Introduction}

The OMC-1S region is a high-luminosity ($\sim 10^4~L_\odot$) source
at infrared and submillimeter wavelengths in the Orion 
Nebula that is still actively forming high-mass stars (Bally, O'Dell, \& McCaughrean
2000). It is located 
about 1' to the SW of the Trapezium
and several powerful molecular outflows (Ziurys, Wilson \& Mauersberger 
1990; Schmid-Burgk et al. 1990; Rodr\'\i guez-Franco, Mart\'\i n-
Pintado \& Wilson 1999) appear to emanate from this
region, probably driven by highly embedded 
young stars (Bachiller et al. 1996). Numerous H$_{2}$O masers (Gaume et al. 1998)
are also known to be present in OMC-1S. 

The survey of proper-motions in the Orion Nebula (Bally et al. 2000;
O'Dell \& Doi 2003)
revealed that at least six relatively large Herbig-Haro outflows 
(HH~202, HH 269, HH 529, HH~203/204, HH~530, 
and possibly HH~528) are originating from a region only a few arcseconds across 
in OMC-1S that 
O'Dell \& Doi (2003) locate at 
$\alpha_{2000}=5^h35^m14^s.56$, $\delta_{2000}=-5^{\circ}$ $23\rlap{'}$ $54\rlap{''}$
\hspace{.1cm} with a positional error of $\pm 1\rlap.{''}5$.      
O'Dell \& Doi (2003) refer to this region as the Optical Outflow Source (OOS).
Although embedded objects (i. e., TPSC-1, Lada et al. 2000) have been 
detected in this region, the actual identification of the
powering sources of the outflows remains uncertain.

From their analysis of 3.6 cm radio continuum VLA archival observations of Orion, 
Zapata et al. (2004) found three faint sources in the OMC-1S
region. Two of them (VLA 13 and VLA 15) are
located near the position for the OOS (and thus may be related to the sources 
that power the multiple outflows that emanate
from this region). The other source (VLA 10) 
is coincident, within the millimeter wavelength positional 
error, with the source FIR 4 (Mezger et al. 1990), the presumed exciting 
source of the low-velocity CO outflow detected by
Schmid-Burgk et al. (1990).   

In this paper we present new observations of the OMC-1S region at 1.3 cm that reveal 
a cluster of eleven sources, of which eight are new
detections in the radio.
This result supports the expectation that
a cluster of young, possibly massive stars should be embedded here. Our results
also allow us to improve
our understanding of the nature of the three sources
previously found at 3.6 cm by Zapata et al. (2004). 

\section{Observations}

The observations were made with the Very Large Array of NRAO\footnote{The National 
Radio Astronomy Observatory is a facility of the National Science Foundation operated
under cooperative agreement by Associated Universities, Inc.} 
in the continuum mode at 1.3-cm during 2003 December 23 and 27. At these epochs,
the VLA was in the B configuration. The phase center of
the observations was $\alpha(2000) = 05^h~ 35^m~ 14\rlap.^s0$;
$\delta(2000) = -05^\circ~ 24'~ 03''$, very close to the
center of the new radio cluster. The absolute amplitude calibrator was 1331+305 and the
phase calibrator was 0541-056, with a bootstrapped flux density of 0.75 $\pm$ 0.01
Jy.

The data were analyzed in the standard manner
using the AIPS package of NRAO and self-calibrated in phase
and amplitude. The images were made using only visibilities
with baselines larger than 90 k$\lambda$, thus suppressing the emission
of structures larger than $\sim$2${''}$.
This procedure is necessary given the presence of complex, extended
emission in the region.
Furthermore, we used the ROBUST parameter of IMAGR set to 0,
in an optimal compromise between sensitivity and
angular resolution. The resulting image rms was 70 $\mu$Jy beam$^{-1}$
at an angular resolution of $0\rlap.{''}30 \times 0\rlap.{''}25$
with PA = $-5^\circ$.
In the OMC-1S region
we detected a total of 11 sources (see Figure 1). 
The positions, flux densities (at 1.3 cm and 3.6 cm), and spectral indices 
of these radio sources
are given in Table 1. The 3.6 cm flux densities are from Zapata et al. (2004).
To refer to these sources we have adopted a position-based nomenclature
after the convention of O'Dell \& Wen (1994). 
In Table 2 we list counterparts (objects within $0\rlap.{''}5$
of the radio source)
for the 1.3 cm sources. Only the radio source 143-353 has an optical
counterpart in the HST optical images of O'Dell \& Doi (2003).

\section{On the Nature of the 1.3 cm Sources}

Following Windhorst et al. (1993), we roughly estimate that the
expected number of 1.3 cm background sources above a flux
density of $S$ is given
by $\langle N \rangle \simeq 1.0 \times 10^{-3}~(S/mJy)^{-1.2}$ $arcmin^{-2}.$
Then, the probability of finding a source with flux density equal or larger than 
0.36 mJy (the weakest source reported) in a $30{''} \times 30{''}$ region is only
$\sim$0.001, so we conclude that most probably all sources are associated
with Orion.
Of the eleven sources detected at 1.3 cm in the field, only three have been
previously detected in the radio at 3.6 cm. The sources 131-411
(= VLA 10), 141-357 (= VLA 13) and 143-353 (= VLA 15) were
detected in the sensitive 3.6 cm image of Zapata et al. (2004).
All the new eight sources have steep positive spectral indices,
which explains why they are detected at 1.3 cm but not at 3.6 cm.
From comparison with the 3.6 cm upper limits we set 
lower limits to their spectral indices, $\alpha > 0.8 \pm 0.3$
(with $S_\nu \propto \nu^{\alpha}$).
This suggests that they may be optically-thick 
H~II regions powered by embedded B-type stars. This 
proposition is consistent with the fact that
three (136-359, 139-357, and 141-357)
of the eight newly found sources are closely associated with the
water vapor masers reported by Gaume et al. (1998).
However, one of the sources (140-410) was found to show large
($\sim$20\% ) right circular polarization. This emission is
believed to trace gyrosynchrotron emission from active magnetospheres
of young, low mass stars. Furthermore, from the results of
Smith et al. (2004) and Robberto et al.
(2004) we find that four other sources (136-356, 136-359, 139-357, 
and 144-351)
are associated with infrared sources
of low bolometric luminosity that cannot drive an H~II region.
We are thus left with the possibility that
some of these new sources are not optically-thick H~II regions,
but may be sources of gyrosynchrotron emission or ionized outflows
from low-mass stars.
We tried unsuccessfully to look for large temporal variations 
by comparing the December 23 and December 27 data.
Observations at several wavelengths and time monitoring are required to
firmly establish the nature of these sources with steep, positive
spectral indices.

\section{Thermal jets}

Only two sources are clearly resolved 
in the 1.3 cm data: 143-353 (= VLA~15), with
deconvolved dimensions of $0\rlap.{''}75 \pm 0\rlap.{''}05 \times
0\rlap.{''}30 \pm 0\rlap.{''}03; PA = 144^\circ \pm 4^\circ$,
and 134-411 (= VLA~10), with
deconvolved dimensions of $0\rlap.{''}28 \pm 0\rlap.{''}02 \times
0\rlap.{''}09 \pm 0\rlap.{''}03; PA = 25^\circ \pm 5^\circ$.
For the other radio sources we derive a typical upper limit 
of $0\rlap.{''}1$ for their angular size.
As we will show, the orientation of these two sources suggests
they are thermal jets that are powering important outflows in the region.

\subsection{143-353}

Within the complex of six major outflows that originate from the OOS
region (O'Dell \& Doi 2003), this feature
aligns with both HH 202 and HH 528. 
HH~202 is the prominent HH object in the central region of the Orion Nebula
and it was one of the first in Orion to be identified as such (Cant\'o et al. 1980;
Meaburn 1986).
Rosado et al. (2001) have proposed that
HH~202 forms a bipolar outflow with HH~203/204,
but the fact that both HH~202 and
HH~203/204 are blueshifted and the excellent alignment
of the 1.3 cm source 143-353, with HH~202 on one side of the radio jet 
and HH~528 on the other side makes this suggestion
less likely. 

Examination of Figure 1 of O'Dell \& Doi (2003)
shows that the PA for HH 202 is $326^\circ \pm 4^\circ$ and that of 
HH 528 is $144^\circ \pm 4^\circ$, i.e. that
they are almost exactly opposite in direction along a common axis
and very well aligned with the major axis of the radio source 143-353
($144^\circ \pm 4^\circ$, see Figure 2).
The proper motion data of O'Dell \& Doi (2003) show that the tip of HH 202 is moving at
51$\pm$13 km s$^{-1}$ towards 
$322^\circ \pm 25^\circ$ and the entire HH 528 object is moving at 25$\pm$14 km s$^{-1}$
towards $159^\circ \pm 35^\circ$, which means that the two 
objects are moving away from one another
and along their axis of orientation.

The radio source 143-353 has a nearly coincident optical counterpart, best seen in
Figure 20 of Bally et al. (2000) as a bright emission line
filament to the northwest from the source marked 9 in that figure. Our source
143-353 is the northwest portion of an irregular filament $3\rlap.{''}2$ long, 
oriented towards $PA=323^\circ \pm 10^\circ$.
The optical source is $0\rlap.{''}8 \times 0\rlap.{''}2$
in angular size, and is visible on HST images in H$\alpha$, [OIII],
[NII], [OI], and [SII]. Its orientation is $PA = 310^\circ \pm 10^\circ$. 
The tangential motion of the
entire filament is 16$\pm$4 km s$^{-1}$ towards $323^\circ \pm 17^\circ$ 
(O'Dell \& Doi 2003)  

It is difficult to assign a source for any of the outflows from the OOS region on
the basis of the extrapolation of orientation vectors as there are multiple 
objects within the error ellipse for the region. However, the additional information
of the optical and radio alignment of 143-353 with both HH 202 and HH 528 suggests
a relationship. In terms of the direction of motion of the optical counterparts
of 143-353, it is plausible that HH 202 is being driven by the source that produces
143-353. The much lower tangential velocity of 143-353 would require that we are actually
seeing the material entrained on the outside of a smaller, higher velocity jet, a condition
that would apply if this jet were passing through the dense layer near the main ionization
front of the nebula.

A connection with HH 528 is suggested by the orientations, but 
143-353 is moving in the opposite direction. However, there may be a common source for
both HH 202 and HH 528 which also produces 143-353. HH 202 is known to have a high
blueshift (the most accurate and recent determination is by Doi, O'Dell \& Hartigan 2004).
A truly bipolar source would mean that HH 528 would have a similar high redshifted 
velocity, for which there is no evidence. However, we do see blueshifted components
from two other (HH 269 and HH 529) opposite moving features coming out of the OOS region 
(Doi, O'Dell \& Hartigan 2004). The east-west orientation of those flows indicates a 
different source within the OOS, but the velocities indicate that some mechanism is
operating that allows opposite flow with common blueshifted components. If this
same mechanism applies to HH 202--143-353--HH528, then they too could be sharing
a common source.

In general, thermal radio jets such as 143-353 are interpreted to mark
the position of the exciting source. 
%The identification of 143-353 as a thermal jet is supported by
%its spectral index of 0.6$\pm$0.1
%(see Table 1), characteristic of ionized outflows.
Indeed, the fact that the source is not optically-thin suggests it is
fairly dense and most probably
directly associated with the exciting star.
However, it is also possible that
143-353 is just part of a long jet, displaced from the exciting star.
If this is the case, there is a good candidate star for producing the 143-353 jet. 
This is the object 145-356, which has been detected at K' 
(Source 9; McCaughrean \& Stauffer
1994), L (Source 2; Lada et al. 2000)
N (Source 60; Robberto et al. 2004),
and three IR bands (Source 3; Smith et al.
2004). There is no radio counterpart to this
source. This apparently stellar source is positioned so that 143-353 lies
$5\rlap.{''}1$ at PA=$307^\circ$. If one accepts the orientation of the extended jet and 143-353 as
definitive, then this star is an excellent candidate for the source. If the optical
proper motions indicate true tangential velocities, then the jet originated
700 years ago. If the proper motion of HH 202 is adopted (which is the better assumption
if the optical material is entrained gas on the edge of the jet), then the age of the
jet is 220 years.

\subsection{134-411}

As mentioned before, three of the eleven 1.3 cm sources had been detected
previously at 3.6 cm (Zapata et al. 2004). One of them,
134-411 (= VLA~10) has a spectral index of 1.9$\pm$0.1, consistent
with an optically-thick H~II region. However, its position
and orientation
suggest that this
source is driving the low-velocity
monopolar molecular outflow (Schmid-Burgk et al. 1990)
that is centered in the 1.3 mm continuum source FIR4 (Mezger et al. 1990),
located at $\alpha(2000) = 05^h~ 35^m~ 13\rlap.^s4;
\delta(2000) = -05^\circ~ 24'~ 13{''}$.
The PA of the major axis of 134-411 (see Figure 3)
is $25^\circ \pm 5^\circ$, while that of the molecular outflow is $31^\circ$
(Schmid-Burgk et al. 1990). 
The association of 134-411 with the outflow is also supported by its spatial coincidence 
with the cluster of high-velocity H$_2$O masers detected and discussed
by Gaume et al. (1998). 

%The remaining two sources that are also detected at 3.6 cm,
%141-357 (= VLA~13) and 143-353 (= VLA 15), have spectral indices
%of 0.6$\pm$0.1, characteristic of ionized outflows. We have noted before
%that 143-353 is angularly resolved and that its major axis aligns with
%two large scale Herbig-Haro flows, HH~202 and HH~528, 
%that emanate from OMC-1S. We then suggest
%that both 141-357 and 143-353 may be powering outflows in the region.
%However, only in the case of 143-353 we resolve the angular
%dimensions of the ionized outflow and can obtain a PA to compare
%with the large scale flows.

\section{Conclusions}

Our results significantly alleviate the apparent lack of exciting sources of the
multiple outflows that originate from OMC-1S. In particular,
the source 143-353 could be driving the HH~202 and possibly the HH~528
flows, while the source 134-411 could be driving the
low-velocity
molecular outflow in the region. However, additional
studies are required to identify the exciting sources of the
multiple outflows and to firmly establish the nature of the objects 
in this new cluster of compact, steep positive spectrum radio sources in Orion.

\vspace{0.5cm}

\acknowledgments
LFR acknowledges the support
of DGAPA, UNAM, and of CONACyT (M\'exico).
This research has made use of the SIMBAD database, 
operated at CDS, Strasbourg, France.

\clearpage
%\pagebreak
\thispagestyle{empty}

\begin{figure}
\centering
\vspace{-1.8cm}
\includegraphics[angle=0,scale=.5]{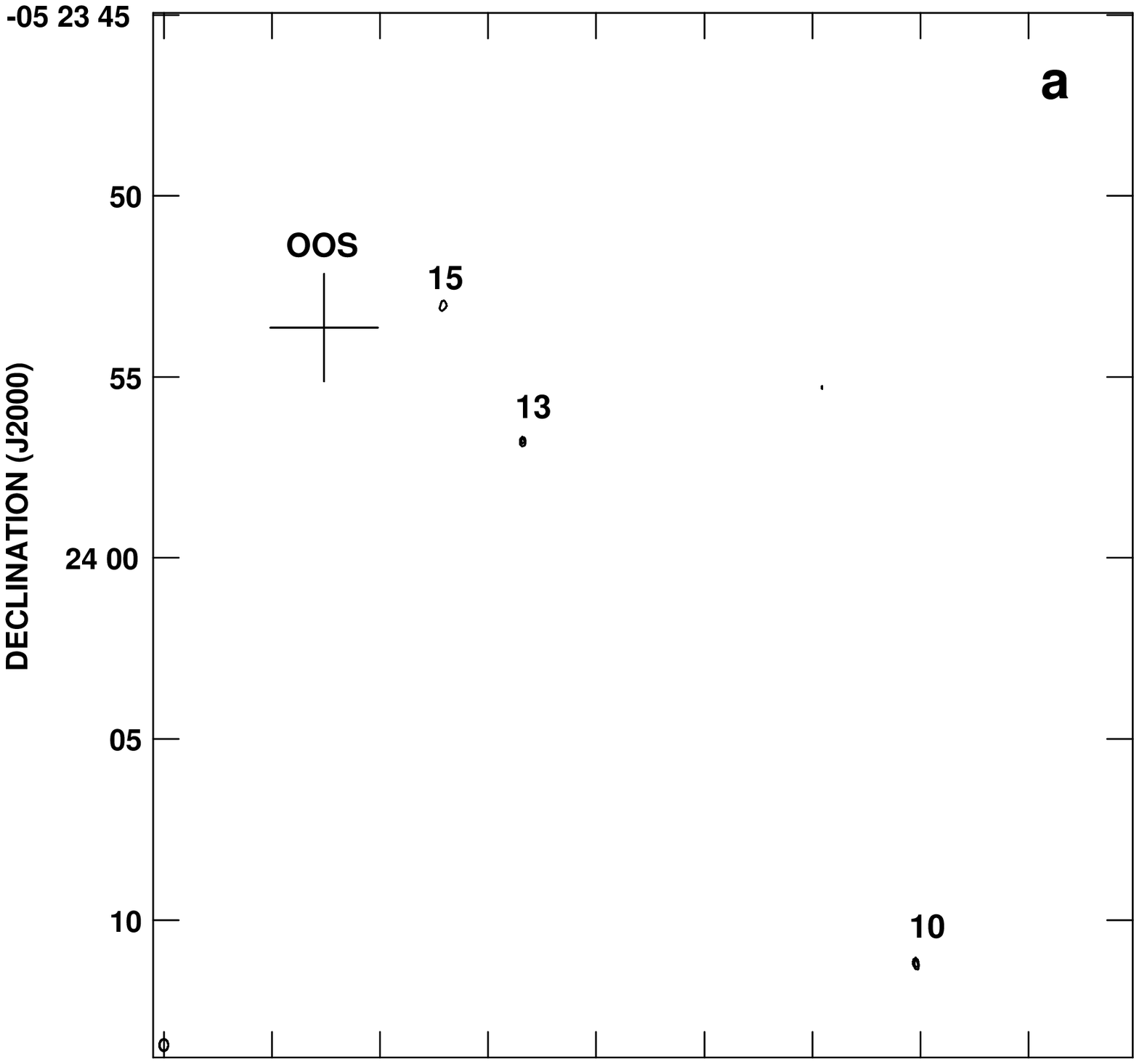}
\vskip-1.8cm
\includegraphics[angle=0,scale=.5]{f1b.eps}
\vskip-1.0cm
\caption{\small VLA continuum images at 3.6 cm and 1.3 cm of the OMC1-S region. 
The position for the OOS (O'Dell and Doi 2003) is indicated with a cross.
The half power contour of the synthesized beams are shown in the bottom left corner
of the images.
{\bf a.} The faint sources 10, 13 and 15 were reported at 3.6 cm by Zapata et al. 2004. 
The contours are -5, 5 and 6 times 50 $\mu$Jy beam$^{-1}$, the rms noise of the image.
{\bf b.} 
The objects 136-359, 136-356, 136-355, 137-347, 139-357, 139-409, 140-410 
and 144-351, are first reported as radio sources here.
The contours are -5, 5, 6, 7, 8, 9, 10, 15, 20, 30 times 70 $\mu$Jy beam$^{-1}$, 
the rms noise of the image.}
\label{fig1}
\end{figure}

\vfill\eject

\begin{figure}
\centering
%\vspace{-2.8cm}
\includegraphics[angle=-90,scale=.6]{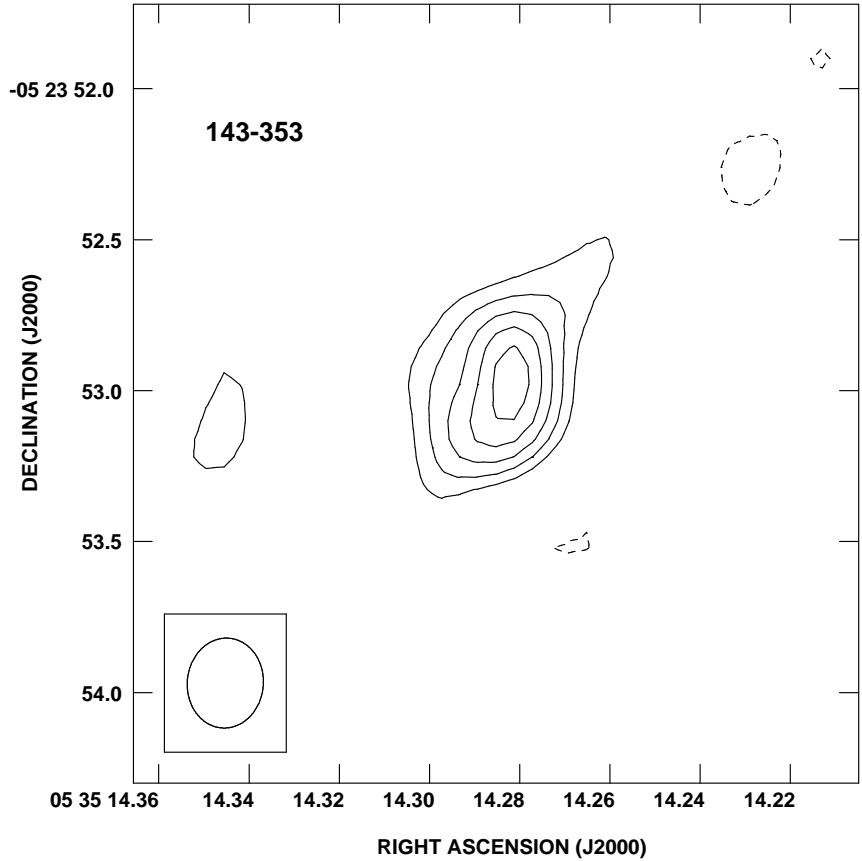}
%\vskip-1.0cm
\caption{\small VLA continuum image at 1.3 cm of the
source 143-353.
The half power contour of the synthesized beam is shown in the bottom left contour.
The contours are -4, -3, 3, 4, 5, 6, and
7 times 70 $\mu$Jy beam$^{-1}$, the rms noise of the image.
This source is proposed to be driving the HH~202 and possibly the HH~528 outflows.
The position angle of HH~202 is $326^\circ \pm 4^\circ$ and that of 
HH 528 is $144^\circ \pm 4^\circ$, aligning well with the position angle
of the major axis of the radio source, which is $144^\circ \pm 4^\circ$.
}
\label{fig2}
\end{figure}

\vfill\eject

\begin{figure}
\centering
%\vspace{-2.8cm}
\includegraphics[angle=-90,scale=.6]{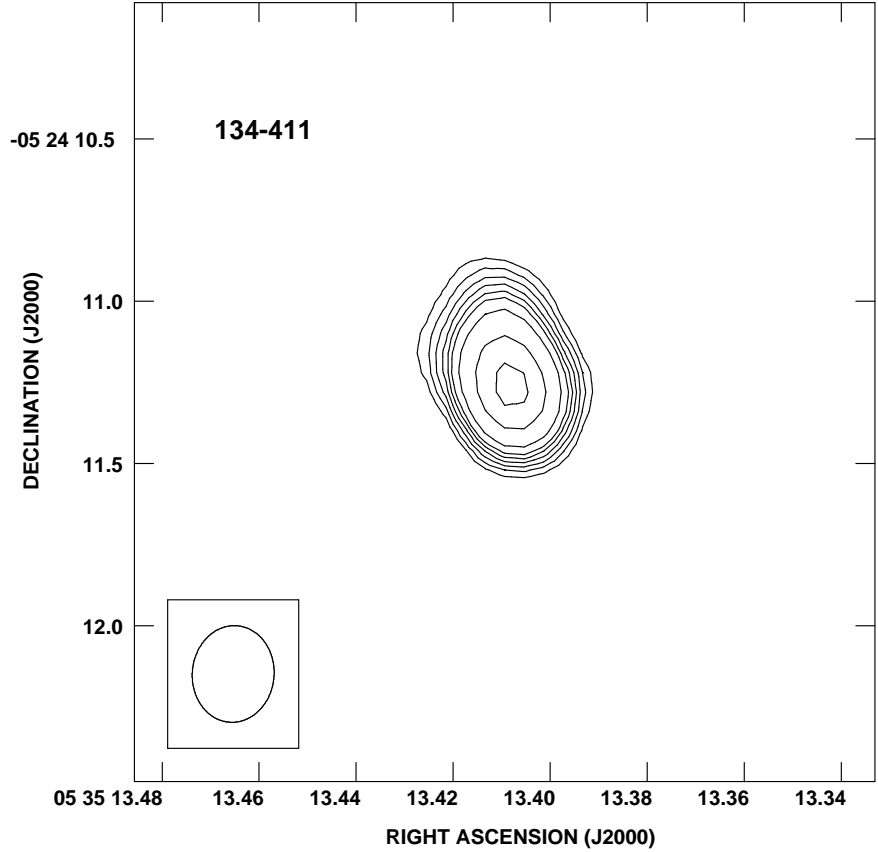}
%\vskip-1.0cm
\caption{\small VLA continuum image at 1.3 cm of the
source 134-411.
The half power contour of the synthesized beam is shown in the bottom left corner.
The contours are -4, -3, 3, 4, 5, 6, 7, 8, 10, 15, and 20
times 70 $\mu$Jy beam$^{-1}$, the rms noise of the image.
This source is proposed to be driving the low-velocity
monopolar molecular outflow associated with the far-infrared source FIR~4.
The position angle of the molecular outflow is $31^\circ$,
in good agreement with the position angle of the major axis of this source,
which is $25^\circ \pm 5^\circ$.
}
\label{fig3}
\end{figure}

\clearpage 
\begin{deluxetable}{l c c c c c c}
%\tabletypesize{\scriptsize}
\tablecaption{Parameters of the VLA Sources Detected Detected at 1.3 cm}
\tablehead{ 
\colhead{}                              &
\colhead{}                              &
\colhead{}                              &
\multicolumn{2}{c}{Flux Density$^a$ } &
\multicolumn{1}{c}{Spectral Index} & \\  
%\multicolumn{3}{c}{Axes$_{1.3 cm}$} \\   
\cline{4-7}                                     
\colhead{Radio}                        &
\colhead{$\alpha_{2000}$}               &
\colhead{$\delta_{2000}$}               &
\colhead{1.3 cm}  &                            
\colhead{3.6 cm}  &
\colhead{1.3 cm/3.6 cm}  & \\
%\colhead{Maj.}  &
%\colhead{Min.}  &
%\colhead{P.A.}  &\\
\colhead{Source}                              &
\colhead{(h m s) }                     &
\colhead{($^\circ$ $^{\prime}$  $^{\prime\prime}$)}              &
\colhead{(mJy)}  & 
\colhead{(mJy)}  &
\colhead{}  &
%\colhead{($^{\prime\prime}$)}  &
%\colhead{($^{\prime\prime}$)}  &
%\colhead{($^\circ$)} &
}
\startdata

134-411  & 05 35 13.408 & -05 24 11.27 &  2.07$\pm$ 0.06  &         0.33 & 1.9 $\pm$0.1        & \\
136-359  & 05 35 13.558 & -05 23 59.06 &  0.93$\pm$ 0.04  &  $\leq$ 0.16 & $\geq$ 1.8$\pm$0.3 & \\
136-356  & 05 35 13.571 & -05 23 55.80 &  0.48$\pm$ 0.04  &  $\leq$ 0.16 & $\geq$ 1.1$\pm$0.3 & \\
136-355  & 05 35 13.636 & -05 23 54.94 &  0.36$\pm$ 0.04  &  $\leq$ 0.16 & $\geq$ 0.8$\pm$0.3 & \\
137-347  & 05 35 13.709 & -05 23 46.89 &  0.76$\pm$ 0.05  &  $\leq$ 0.16 & $\geq$ 1.6$\pm$0.3 & \\
139-357  & 05 35 13.875 & -05 23 57.21 &  0.67$\pm$ 0.04  &  $\leq$ 0.16 & $\geq$ 1.4$\pm$0.3 & \\
139-409  & 05 35 13.930 & -05 24 09.43 &  1.03$\pm$ 0.06  &  $\leq$ 0.16 & $\geq$ 1.9$\pm$0.2 & \\
140-410  & 05 35 13.972 & -05 24 09.84 &  1.63$\pm$ 0.06  &  $\leq$ 0.16 & $\geq$ 2.4$\pm$0.3 & \\
141-357  & 05 35 14.135 & -05 23 56.69 &  0.90$\pm$ 0.04  &         0.32 &  1.1$\pm$0.1         & \\
143-353  & 05 35 14.281 & -05 23 52.93 &  1.60$\pm$ 0.06  &         0.30 &  1.7$\pm$0.1         & \\
144-351  & 05 35 14.391 & -05 23 50.81 &  1.01$\pm$ 0.05  &  $\leq$ 0.16 & $\geq$ 1.9$\pm$0.3   &\\

\enddata
\tablecomments{
                (a): Total flux density corrected for primary beam response.}
\end{deluxetable}

\clearpage
\begin{deluxetable}{l c }
%\tabletypesize{\scriptsize}
\tablecaption{Counterparts of the 1.3 cm VLA Sources}
\tablehead{
\colhead{Source}                       & 
\colhead{Counterpart$^a$ } \\
}
\startdata
134-411 & VLA 10, GWV, FIR 4 \\
136-359 & F 370, GWV, SBSMH 4, MAX 42 \\
136-356 & SCH 18, TPSC 46, SBSMH 5, MAX 43 \\
136-355 & -\\
137-347 & AD 2650\\
139-357 & GWV, SBSMH 6\\
139-409 & GWV\\
140-410 & -\\
141-357 & GWV, VLA 13\\
143-353 & VLA 15, HST 143-353 \\
144-351 & SBSMH 2, MAX 58\\        

\enddata
\tablecomments{
                (a): Sources at other wavelengths within
   $0\rlap.{''}5$ of the 1.3 cm sources. AD = Ali \& Depoy 1995 (K-band);
                %BPN78 = Beckwith 1978(K band); 
                %BSD98 = Bally et al. 1998 (UV);
                %CHS01 = Carpenter et al. 2001 (K-band);
%                GSL02 = Giveon et al. 2002 (regions HII);
                %D93  =  Dougados et al. (1993) (L-band);
%               GCS95 = Gagne et al. 1995 (X-rays);
                F = Feigelson et al. 2002 (X-rays);
                FIR = Mezger et al. 1990;
%               GMR = Garay, Moran, \& Reid 1987 (radio centimeter);
                GWV = Gaume et al. 1998 (water masers);
                %HAB = Hyland et al. 84 (K band); 
                %Hb97 = Hillenbrand 1997 (I-band);
                %HC = Hillenbrand \& Carpenter 2000 (K-band);
                %LRY00 = Luhman et al. 2000 (K-band);
                %MLL95 = Mundy et al. 1995 (radio millimeter);
                %OW94 = O'Dell \& Wen 1994 (visible);
		HST = O'Dell \& Doi 2003 (visible)
		MAX = Robberto et al. 2004 (10 and 20 $\mu$m)
                SBSMH = Smith et al. 2004 (8.8 and 11.7 $\mu$m)
                SCH = Schulz et al. 2001 (X-rays);
                %SCB99 = Simon et al. 1999 (K-band).
                VLA = Zapata et al. 2004 (radio 3.6 cm)
%               TPSC = Lada et al. 2000 (I-band) 
                \\}
\end{deluxetable}

\end{document}